\documentclass[lettersize,journal]{IEEEtran}
\usepackage{amsmath,amsfonts}
\usepackage{algorithmic}
\usepackage{algorithm}
\usepackage{array}
\usepackage[caption=false,font=normalsize,labelfont=sf,textfont=sf]{subfig}
\usepackage{textcomp}
\usepackage{stfloats}
\usepackage{url}
\usepackage{verbatim}
\usepackage{graphicx}
\usepackage{cite}
\usepackage{circuitikz}
\usepackage{multicol}
\begin{document}

%MB add
\onecolumn

\title{Representation of a Noisy Transmission Line}

\author{Martin Bucher and Daniel Molnar\\
\today}

\markboth{Journal of \LaTeX\ Class Files,~Vol.~14, No.~8, August~2021}%
{Shell \MakeLowercase{\textit{et al.}}: A Sample Article Using IEEEtran.cls for IEEE Journals}
\IEEEpubid{0000--0000/00\$00.00~\copyright~2021 IEEE}

\maketitle

%MB add
\begin{multicols}{2}
%\end{multicols}{2}

\begin{abstract}
\bf 
\noindent 
We analyse a lossy transmission line and the Johnson-Nyquist 
noise generated therein.
A representation as a noisy two-port with a voltage and a 
current noise sources on one 
end of a noiseless two-port is given. An expression for the 
noise properties
is given for an arbitrary temperature profile along the 
transmission line. Compatibility of the general expression 
found here
with expectations for special cases calculable using thermodynamics
is demonstrated.
\end{abstract}


\begin{figure}[H]%[!h]
\centering
\includegraphics[width=2.5in]{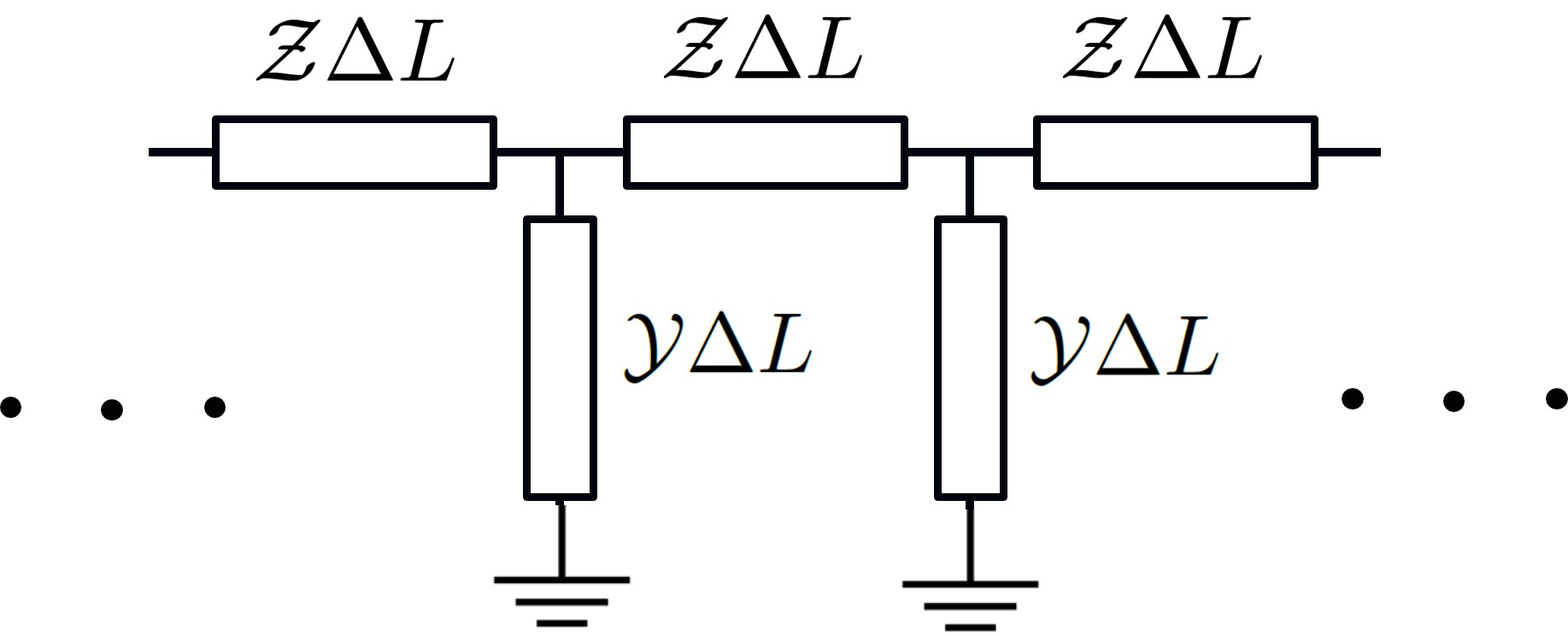}
\caption{A cable of uniform along its length is represented as an alternating sequence of
series and shunt elements chained together.}
\label{fig_five}
\end{figure}

\section{Motivation}

Here we investigate how to model and to predict the noise generated in a lossy 
transmission line at high accuracy and in the case where the temperature is not
uniform along the cable length. 
For many applications the level of detail developed in this paper is not needed.
Our motivation is to better model noise in radio astronomy experiments that aim to measure
the so-called 21 cm global signal. 
(See for example \cite{global21cmTheoryReview} and references 
therein for a discussion.)
These experiments seek to measure predicted deviations of order a few tens of mK from the perfect 2.75 K CMB (Cosmic Microwave Background),
which must be separated from synchrotron emission from our own Galaxy
(separable on account of its smooth frequency dependence) 
whose Rayleigh-Jeans brightness temperature 
at the frequencies of interest 
is of order $10^{3}$--$10^{4}$ K around the 
Galactic plane. 
In contrast to measurements of the CMB anisotropies, where
the data consist of measurements of small temperature differences 
between different directions in the sky, global 21 cm experiments have more the character of an absolute measurement of electrical noise. A detection of a 21cm absorption trough 
centered around 78 MHz deeper than the theoretically expected
has been claimed \cite{edgesDetectionPaper}, 
but it is presently unclear whether
this feature is a genuine sky signal or 
an instrumental artefact.  Other experiments are underway
to verify this claim and detect the elusive 21 cm global signal. (See for example \cite{reachNaturePaper} and references 
therein.) 
The case will remain open until
a convincing argument can be put forward that the instrumental setup
has been correctly modelled at the requisite accuracy.

Since the amplifier chain cannot be calibrated absolutely, thermal (Johnson-Nyquist)
noise sources (i.e., resistors at  known temperatures) and calibrated noise sources
(e.g., standard noise diodes) having various source impedances are used to calibrate
the amplifier gain  \cite{rogersBowmanPaper,rogersMemo,edgesCalibration,reachCalPaper}
and measure the four real noise parameters of the LNA \cite{reprNoisyTwoPorts}. In order 
to characterize these quantities at the required accuracy, the contribution of
the Johnson-Nyquist noise generated within the connecting transmission lines (i.e., cables) due to resistive losses needs to be calculated.

In this paper, we consider only Johnson-Nyquist noise, 
and linearity is always assumed. 
We note that if the conductors of the transmission line also 
carry a DC bias or power supply current, additional noise may be generated
due to resistance fluctuations, and this excess noise adds in quadrature.
Such excess noise whose power is proportional to the square of the DC current
typically has a red spectrum, with $P\sim \nu ^{-\alpha }$ where $\alpha \approx 1.$
The pure power law here is only an approximation:
unlike for the case of Johnson-Nyquist noise, resistance 
fluctuations do not follow a precise universal functional form.
Nothing universal may be said about such
excess noise, and we are not considering this type of noise in the current work. 
We also assume the absence of long-range spatial correlations in the 
Johnson-Nyquist noise.

\section{General Representation of a Cable}

A cable may be modelled as an infinite sequence of infinitesimal series and shunt elements 
chained together,
as illustrated in Fig.~\ref{fig_five}.
These may most conveniently be represented in the transfer matrix (or
$ABCD$-matrix) representation (see e.g., \cite{pozar}). From the representation 
of a cable element of infinitesimal length, 
the transfer matrix for a cable of 
finite length $L$ can be derived using the usual rules of matrix exponentiation. 
We write 
\begin{equation}
\begin{aligned}
\mathcal{T}_{series}(\Delta L)
&=
\begin{pmatrix}
1 & 
\mathcal{Z}(\Delta L)
\\
0 & 1
\end{pmatrix} +O(\Delta L^2)\\[6pt]
&=
\begin{pmatrix}
1 & 
+j\bar{\mathcal{X}}(\Delta L)\\
0 & 1
\end{pmatrix} +O(\Delta L^2),
\\[6pt]
\mathcal{T}_{shunt}(\Delta L)&=
\begin{pmatrix}
1 & 0 \\
\mathcal{Y}(\Delta L)
 & 1
\end{pmatrix} +O(\Delta L^2)
\\[6pt]
&=
\begin{pmatrix}
1 & 0 \\
+j\bar {\mathcal{B}}(\Delta L)
 & 1
\end{pmatrix} +O(\Delta L^2)
\end{aligned}
\end{equation}
where $\mathcal{Z}$
and $\mathcal{Y}$ are impedance and admittance per unit length, respectively.
Here we define  complex versions of the 
reactance and susceptance per unit length
as
$\bar {\mathcal{X}}= \mathcal{X}-j\mathcal{R}$
and 
$\bar {\mathcal{B}}= \mathcal{B}-j\mathcal{G}$
in order the facilitate comparison with lossless
case.
We work in a narrow frequency band so the frequency dependence is implicit. Such a representation is always possible when the cable respects reciprocity, in which case a 2-port
always has a $\pi $ or $T$ representation. Homogeneity along the length of the cable is also assumed, as is linearity.
For a finite length $L$ of cable, we obtain the transfer matrix
\end{multicols}
\hrule 

\vskip 10pt 

\begin{equation}
\begin{aligned}
\mathcal{T}(L)&= 
\exp \left[
\begin{pmatrix}
0 & 
j\bar{\mathcal{X}}\\
j\bar{\mathcal{B}}
 & 0
\end{pmatrix} L
\right]\\
&=
\begin{pmatrix}
1 & 0\\ 
0 & 
\sqrt{
\bar{\mathcal{B}}/
\bar{\mathcal{X}}
}
\end{pmatrix} 
\exp \left[+j
\sqrt{
\bar{\mathcal{B}}
\bar{\mathcal{X}}
}L\sigma _1
\right]
\begin{pmatrix}
1 & 0\\ 
0 & 
\sqrt{
\bar{\mathcal{X}}/
\bar{\mathcal{B}}
}
\end{pmatrix} 
\\ &=
\begin{pmatrix}
1 & 0\\ 
0 & 
\sqrt{
\bar{\mathcal{B}}/
\bar{\mathcal{X}}
}
\end{pmatrix} 
\Bigl[
\cos \bigl[
(k-j\gamma )
L\bigr]\sigma _0
+j\sin \bigl[
(k-j\gamma )
L\bigr] \sigma _1
\Bigr]
\begin{pmatrix}
1 & 0\\ 
0 & 
\sqrt{
\bar{\mathcal{X}}/
\bar{\mathcal{B}}
}
\end{pmatrix} 
\\
&=
\begin{pmatrix}
1 & 0\\ 
0 & 
\sqrt{
\bar{\mathcal{B}}/
\bar{\mathcal{X}}
}
\end{pmatrix} 
%\Bigl[
\left[
\frac{
e^{+jkL+\gamma L}
+
e^{-jkL-\gamma L}
}{2}
\sigma _0
+
\frac{
e^{+jkL+\gamma L}
-
e^{-jkL-\gamma L}
}{2}
\sigma _1
\right ]
%\Bigr]
%\\ &\qquad \times
\begin{pmatrix}
1 & 0\\ 
0 & 
\sqrt{
\bar{\mathcal{X}}/
\bar{\mathcal{B}}
}
\end{pmatrix} 
\end{aligned}
\end{equation}
\vskip 2pt 
\hrule 
\vskip 2pt 
\begin{multicols}{2}
\noindent 
where 
$\sqrt{
\bar{\mathcal{B}}
\bar{\mathcal{X}}
}=\bar k=k-j\gamma .$
We may rewrite
\begin{equation}
\mathcal{T}=
\begin{pmatrix}
\phantom{j{Z_c}^{-1}}
\cos (\bar kL)
&
+j{Z_c}
\sin (\bar kL)
\\
+j{Z_c}^{-1}
\sin (\bar kL)
&
\phantom{j{Z_c}}
\cos (\bar kL)
\end{pmatrix}. 
\end{equation}
In terms of the elements of the transfer matrix, the impedance
matrix $\mathbf{Z}$ is given by
\begin{equation}
\mathbf{Z}
=
\begin{pmatrix}
\dfrac{
A
}{
C
} &
\dfrac{
AD-BC
}{C
} \\[3mm]
\dfrac{1
}{C
} &
\dfrac{D
}{C
}
\end{pmatrix}, 
\end{equation}
so that for the above
\begin{equation}
\mathbf{Z}
=
Z_c
\begin{pmatrix}
-j\cot (\bar kL) &
-j\csc (\bar kL)
\\
-j\csc (\bar kL)
&
-j\cot (\bar kL)
\end{pmatrix}.
\end{equation}
For an infinitely long cable, as $L\to \infty ,$
$\cot (\bar kL) \to +j$
and
$\cot (\bar kL) \to +j$
because $\textrm{Im}(\bar k)<0.$
Consequently, in this limit 
\begin{equation}
\mathbf{Z}
\to 
\begin{pmatrix}
Z_c & 0\cr
0 & Z_c
\end{pmatrix}.
\end{equation}
Note that $Z_c$ in the general case has a non-vanishing imaginary
part, so that no reflectionless matching with a lossless transmission
line is possible.

For a finite section of cable it is useful to express the cable behavior in terms of an
$S$-matrix with respect to some real reference characteristic impedance $\bar Z_c$
(e.g., $\bar Z_c=50~\Omega ),$ so that
\begin{equation}
\mathbf{S}
=
\frac{
\mathbf{Z}-\bar Z_c\mathbf{I}
}{
\mathbf{Z}+\bar Z_c\mathbf{I}
}.
\end{equation}
Here $Z_c=
\sqrt{ \mathcal{Z}/ \mathcal{Y}}
=
\sqrt{ \bar{\mathcal{X}}/ \bar{\mathcal{B}}}$ 
acts as a characteristic impedance of the lossy transmission line, which 
in the lossy case is not real. We see that the cable is characterized by four real 
parameters. Any symmetric reciprocal
cable is completely characterized by the above parameterization. 

%$\mathbf{1}$

%The above derivation for the cable transfer matrix is assumed to consist of uniform bits characterized by $\mathcal{Z}, \mathcal{Y}$. 
%If we assume that uniformity is not satisfied, $\mathcal{T}$ becomes a local density type quantity. Thereby, we can employ similar ideas. The cable in this case may have varying parameters along its length, notably the complex propagation constant will be a function of length now. In which case the total length transfer matrix $\mathcal{T_L}
%$ is 

%\begin{equation}
%\prod_{i=1}^{N} \mathcal{T}_i(l_i)
%=
%\sum log\mathcal{T}_i(l_i)
%\end{equation}

%this formula already presents a numerical method for evaluation if the cable has some known number of discrete pieces having different transfer matrices, $\mathcal{T}_i(l_i)$. This could simply be a consequence of a composite cable or other non-uniformity along its length.  This idea can be extended to give an ABCD matrix density which is a function of length, that we can call $\mathcal{A}$. 

%\begin{equation}
%\mathcal{T}_L
%=
%\int \mathcal{A} dl
%\end{equation}

%This can be especially useful for the so-called forward modelling in which case a cable can be built up from any arbitrary function. This is also helpful for the reverse modelling for we can build up a model for both physical temperature and charateristic impedance that is varying along the cable length that can be fit to measurement data. We mention this as an aside, as it bears some practical relevance for the applications of radioastronomy receivers which is our primary motivation throughout this work. 

Unlike in the time domain Telegrapher's equation where the series and 
shunt resistances are independent of frequency (as are the series inductance and shunt capacitance), here no assumption is 
made that the propagation can be adequately modelled 
by a PDE in $z$ and $t.$ In reality, as a consequence of the 
skin effect, the effective series resistance increases with 
frequency, and the attenuation due to dielectric loss 
(represented as shunt resistance) increases even more rapidly 
with frequency.
Here no particular dependence on frequency is assumed for the
effective series impedance and shunt admittance. 

\section{The Noisy Cable} 

The description above is so far noiseless. We have not 
yet associated  Johnson-Nyquist noise with the resistive 
elements. Assuming a physical temperature $T,$ we replace each series 
resistor in the infinitesimal equivalent circuit
with a resistor and a random voltage source in series, and 
similarly we replace each shunt resistor with the same shunt 
resistor and a random current source in parallel (Th\'evenin and Norton equivalent). 

This is not a convenient representation, as in general we would wish to view 
the cable as a noisy two-port, 
moving all the distributed noise sources to the output end of 
the cable. How to move noise sources between ports 
for the general case of an N-port is described in detail in Ref.~\cite{bucher}
This can be done simply using the form of the 
transfer matrix found above, so that
\begin{equation}
\begin{aligned}
&\begin{pmatrix}
v_{out}(L)
\\
i_{out}(L)
\end{pmatrix}
=
\int _0^L\!\!
dL'~\\
&\qquad \times 
\begin{pmatrix}
T_{vv}(L_L')
&
T_{vi}(L_L')
\\[6pt]
T_{iv}(L_L')
&
T_{ii}(L_L')
\end{pmatrix}
\begin{pmatrix}
\frac{\delta v(L')}{dL'}
\\
\frac{\delta i(L')}{dL'}
\end{pmatrix}.
\end{aligned}
\end{equation}
Here $\delta v(L')$ and $\delta i(L')$ are the series voltage
and shunt current sources 
associated with the interval $[L', L'+dL'].$ So far these are 
infinitesimal complex numbers.
We have not so far made any assumptions regarding how these are 
determined (e.g., from Johnson-Nyquist noise). 
For the correlation functions, we obtain 
\end{multicols}
\vskip -6pt 
\hrule
\vskip 2pt 
\begin{equation}
\begin{aligned}
&
{\setlength\arraycolsep{2pt}
\begin{pmatrix}
\left\langle v_{out}(L) v_{out}(L)^* \right\rangle
&
\left\langle v_{out}(L) i_{out}(L)^* \right\rangle
\\[3pt]
\left\langle i_{out}(L) v_{out}(L)^* \right\rangle
&
\left\langle i_{out}(L) i_{out}(L)^* \right\rangle
\end{pmatrix}
}
=
\int _0^L\!\!dL' 
\int _0^L\!\!dL^{\prime \prime }
\mathbf{T}(L;L')
%\\ &\qquad \times 
{\setlength\arraycolsep{2pt}
\begin{pmatrix}
\left\langle \delta v(L') \delta v(L^{\prime \prime })^* \right\rangle
&
\left\langle \delta v(L') \delta i(L^{\prime \prime })^* \right\rangle
\\[3pt]
\left\langle \delta i(L') \delta v(L^{\prime \prime })^* \right\rangle
&
\left\langle \delta i(L') \delta i(L^{\prime \prime })^* \right\rangle
\end{pmatrix}
}
\mathbf{T}^\dagger (L;L^{\prime \prime })
\end{aligned}
\label{eqn:InhomoCable}
\end{equation}
\vskip 2pt 
\hrule

\vskip -2pt 

\begin{multicols}{2}
In the case of spatially white noise 
(which here follows from the hypothesis of translation
invariance),
this simplifies to
\begin{equation}
\begin{aligned}
&
\begin{pmatrix}
\left\langle v_{out}(L) v_{out}(L)^* \right\rangle
&
\left\langle v_{out}(L) i_{out}(L)^* \right\rangle
\\[2pt]
\left\langle i_{out}(L) v_{out}(L)^* \right\rangle
&
\left\langle i_{out}(L) i_{out}(L)^* \right\rangle
\end{pmatrix}
\\ 
&
=
\int _0^LdL' 
\mathbf{T}(L;L')
\begin{pmatrix}
W_{vv}(L')
&
W_{vi}(L')
\\
W_{iv}(L')
&
W_{ii}(L')
\end{pmatrix}
\mathbf{T}^\dagger (L;L^{\prime })
\end{aligned}
\end{equation}
where 
\begin{equation}
\left\langle \delta a(L') ~\delta b(L^{\prime \prime })^* \right\rangle
=\delta (L'-L^{\prime \prime })~W_{ab}(L').
\end{equation}

Assuming Johnson noise in the resistive elements of the infinitesimal equivalent circuit for bandwidth $B$
and in the absence of correlation in the noise of the shunt and series resistances, we obtain
\begin{equation}
\begin{aligned}
W_{vv}&=4\mathcal {R}k_BT(L)B,\\
W_{ii}&=4\mathcal {G}k_BT(L)B,\\
W_{vi}&=
W_{iv}=0.
\end{aligned}
\end{equation}
The cross term vanishes by reflection symmetry about the cable cross section. 
Here we have written $T(L)$ in the place of $T$ in order to admit the possibility of a varying temperature along the cable.
\end{multicols}
\vskip 2pt
\hrule 
\vskip 6pt
\begin{equation}
{%\small 
\begin{aligned}
&\begin{pmatrix}
\left\langle v_{out}(L) v_{out}(L)^* \right\rangle
&
\left\langle v_{out}(L) i_{out}(L)^* \right\rangle
\\[3pt]
\left\langle i_{out}(L) v_{out}(L)^* \right\rangle
&
\left\langle i_{out}(L) i_{out}(L)^* \right\rangle
\end{pmatrix}
=4k_BB\int _0^Ldx ~ T(x) 
\\[7pt]
&\qquad \qquad \times
\begin{pmatrix}
 \cos [\bar kx]
&
jZ_c\sin [\bar kx]
\\
j\sin [\bar kx]/Z_c 
&
 \cos [\bar kx]
\end{pmatrix}
\begin{pmatrix}
\mathcal{R} & 0           \\
0           & \mathcal{G}
\end{pmatrix}
\begin{pmatrix}
 \cos [\bar k^*x]
&
-j\sin [\bar kx^*x]/Z_c^*
\\
-j Z_c^* \sin [\bar k^*x]
&
 \cos [\bar k^*x]
\end{pmatrix}
\\[7pt]
&=4k_BB\int _0^Ldx ~ T(x) \left[
\mathcal{R}
\begin{pmatrix}
\phantom{+j}
\cos [\bar kx] \cos [\bar k^*x]
&
-j\dfrac{
\cos [\bar kx] \sin [\bar k^*x]
}{
Z_c^*
}
\\[7pt]
+j\dfrac{
\sin [\bar kx] \cos [\bar k^*x]
}{
Z_c
}
&
\phantom{-j}
\dfrac{
\sin [\bar kx] \sin [\bar k^*x]
}{
Z_c
Z_c^*
}
\end{pmatrix}
\right.
\\[7pt]
&\qquad \qquad+
\mathcal{G}
\left.
\begin{pmatrix}
Z_cZ_c^*
\sin [\bar kx] \sin [\bar k^*x]
&
+jZ_c \sin [\bar kx] \cos [\bar k^*x]
\\[10pt]
-jZ_c^*\cos [\bar kx] \sin  [\bar k^*x]
&
+
\cos [\bar kx] \cos [\bar k^*x]
\end{pmatrix}
\right] 
\end{aligned}
}
\label{eqn:noiseCorr}
\end{equation}
%\vskip 2pt
%\hrule 
%\vskip 2pt
%\begin{multicols}{2}

The above general formula can be evaluated  analytically 
for a uniform physical temperature along the length of cable 
as well as for some simple $T(L)$ profiles.
Otherwise numerical integration is required.
In the case of uniform physical temperature along the cable's length, 
i.e., $T(L)=T_0,$  
%eqn.~(15) 
eqn.(\ref{eqn:noiseCorr})
can be evaluated analytically to give
%\end{multicols}
%\vskip 2pt
%\hrule 
%\vskip 2pt
\begin{equation}
{\small 
\begin{aligned}
&\begin{pmatrix}
\left\langle v_{out}(L) v_{out}(L)^* \right\rangle
&
\left\langle v_{out}(L) i_{out}(L)^* \right\rangle
\\[9pt]
\left\langle i_{out}(L) v_{out}(L)^* \right\rangle
&
\left\langle i_{out}(L) i_{out}(L)^* \right\rangle
\end{pmatrix}
=
4
\mathcal{R}
k_BT_0B 
\begin{pmatrix}
\left[
\dfrac{\sinh (2\gamma L)}{4\gamma }
+
\dfrac{\sin (2kL)}{4k}
\right]
& 
\dfrac{1}{Z_c^*}
\left[
+\dfrac{\sinh ^2(\gamma L)}{2\gamma }
-\dfrac{j\sin ^2(k L)}{2k}
\right]
\\[9pt]
\dfrac{1}{Z_c}
\left[
+\dfrac{\sinh ^2(\gamma L)}{2\gamma }
+\dfrac{j\sin ^2(k L)}{2k}
\right]
& 
\dfrac{1}{
Z_cZ_c^*}
\left[
\dfrac{\sinh (2\gamma L)}{4\gamma }
-
\dfrac{\sin (2kL)}{4k}
\right]
\end{pmatrix}
\\[9pt]
&~~+4
\mathcal{G}
k_BT_0B
\begin{pmatrix}
Z_cZ_c^*
\left[
\dfrac{\sinh (2\gamma L)}{4\gamma }
-
\dfrac{\sin (2kL)}{4k}
\right]
& 
Z_c
\left[
+\dfrac{\sinh ^2(\gamma L)}{2\gamma }
+\dfrac{j\sin ^2(k L)}{2k}
\right]
 \\[9pt]
Z_c^*
\left[
+\dfrac{\sinh ^2(\gamma L)}{2\gamma }
-\dfrac{j\sin ^2(k L)}{2k}
\right]
& 
\left[
\dfrac{\sinh (2\gamma L)}{4\gamma }
+
\dfrac{\sin (2kL)}{4k}
\right]
\end{pmatrix}.
\end{aligned}
}
\label{eqn:CorrUnifTemp}
\end{equation}
The correlation matrix in eqn.~(\ref{eqn:CorrUnifTemp})
may be transformed to the travelling wave representation 
\cite{kurokawa,meysPaper}
with respect to a reference characteristic impedance
$Z_{ref}$ where $v_{L,R}=(v\mp Z_{ref}i)/2\sqrt{\textrm{Re}(Z_{ref})}.$ Setting $Z_{ref}=Z_c,$ we obtain
\begin{equation}
\begin{aligned}
\begin{pmatrix}
\left\langle v_L(L) v_L(L)^* \right\rangle
&
\left\langle v_L(L) v_R(L)^* \right\rangle
\\[9pt]
\left\langle v_R(L) v_L(L)^* \right\rangle
&
\left\langle v_R(L) v_R(L)^* \right\rangle
\end{pmatrix}
&=
\frac{
4\mathcal{R}k_BT_0B }{4\textrm{Re}(Z_{ref})}
\begin{pmatrix}
\dfrac{1-\exp (-2\gamma L)}{2\gamma } &
\dfrac{-j(\exp (-2jkL)-1)}{2k}\\
\dfrac{+j(\exp (+2jkL)-1)}{2k}& 
\dfrac{\exp (+2\gamma L)-1}{2\gamma }
\end{pmatrix}
\\
& \qquad
+
\frac{4\mathcal{G}k_BT_0B~\vert Z_c\vert ^2}
{4\textrm{Re}(Z_{ref})}
\begin{pmatrix}
\dfrac{1-\exp (-2\gamma L)}{2\gamma } &
\dfrac{-j(\exp (-2jkL)-1)}{2k}\\
\dfrac{+j(\exp (+2jkL)-1)}{2k}& 
\dfrac{\exp (+2\gamma L)-1}{2\gamma }
\end{pmatrix}.
\end{aligned}
\label{eqn:CorrUnifTempTW}
\end{equation}

\vskip 6pt
\hrule 
\vskip 6pt

%COMMENTARY ABOUT THE TRAVELLING WAVE CORRELATION HERE.

%\begin{multicols}{2}
%Furthermore, this can be rewritten in a convenient form for a f
%our noise 
%parameter representation from the noise-correlation matrix form if we 
%make the following identifications: 
%\end{multicols}
%\vskip 2pt
%\hrule 
%\vskip 2pt
%\begin{equation}
%{\small 
%\begin{aligned}
%&\begin{pmatrix}
%\left\langle v_{out}(L) v_{out}(L)^* \right\rangle
%&
%\left\langle v_{out}(L) i_{out}(L)^* \right\rangle
%\\[5pt]
%\left\langle i_{out}(L) v_{out}(L)^* \right\rangle
%&
%\left\langle i_{out}(L) i_{out}(L)^* \right\rangle
%\end{pmatrix}
%=
%4k_BT_0
%\begin{pmatrix}
%R_n
%& 
%\dfrac{F_{min}-1}{2}-R_nY_{opt}^*
% \\
%\dfrac{F_{min}-1}{2}-R_nY_{opt}
%& 
%R_n|Y_{opt}|^2
%\end{pmatrix}
%\end{aligned}
%}
%\end{equation}
%\vskip 2pt
%\hrule 
%\vskip 2pt

\begin{multicols}{2}
\noindent
We observe that as the cable becomes long, the amplitude of the travelling noise wave
propagating into the cable diverges while the noise wave emanating from the cable
approaches a constant amplitude. This entering divergent wave is needed to produce
the wave incident on the other end of the cable. The divergence is exactly such 
that this wave has an amplitude approaching a constant by the time it reaches the other end
at $x=L$ when $L$ is large.

%For a cable many decay lengths long (i.e., $\gamma L \gg 1$), 
%the condition number of the correlationnmatrix becomes
%very large, as does the absolute value of the matrix elements.
%This is because there is a part singular in the $L\to \infty $
%limit proportional to the zero rank matrix
%\begin{equation}
%\begin{pmatrix}
%1 
%\\
%-Z_c^{-1}
%\end{pmatrix}
%%\otimes
%\begin{pmatrix}
%1 
%&
%-Z_c^{-1*}
%\end{pmatrix}
%=
%\begin{pmatrix}
%1
%&
%-Z_c^{-1*}
%\\
%-Z_c^{-1}
%&
%(Z_cZ_c^*)^{-1}
%\end{pmatrix}
%\end{equation}
%which is an outer product of two vectors expressed in matrix notation. 
%This singular part corresponds to a completely correlated pair of voltage and current sources
%whose relative amplitudes are 
%tuned so that in the case of the infinitely long cable, a 
%current flows into the cable and is completely absorbed
%with no reflection. 
%For a finite cable, however, this term does create the right
%behavior at the opposite extremity of the cable. Because of the 
%large attenuation, or 
%near perfect isolation of the two extremities, this component 
%must diverge as the cable length
%is taken to become large. However there is no divergence as 
%seen by other ports connected to the cable. 

\section{Compatibility With Thermodynamic Expectations}

In the above discussion, a section of lossy cable at temperature 
$T$ was represented as a noisy
2-port whose noise properties were inferred by replacing each resistor
in the infinitesimal equivalent circuit with the resistor-source combination
suggested by the Johnson-Nyquist formula \cite{johnson,nyquist,solidStateNoiseBook}.
Here we use as a starting point the S-matrix representation for a finite cable
section and derive the constraints imposed by considerations of thermodynamic
equilibrium. For the sequence of ports
$\textrm{(one-port)}_1$-$\textrm{(two-port)}$-$\textrm{(one-port)}_2$
connected together and all at the same temperature, the noise sources
must have values such that there is no net power transfer across the two junction.
We verify that the cable model with noise sources as computed in the previous section
satisfies these conditions. 

\begin{figure}[H]%[!h]
\centering
\includegraphics[scale=0.5]{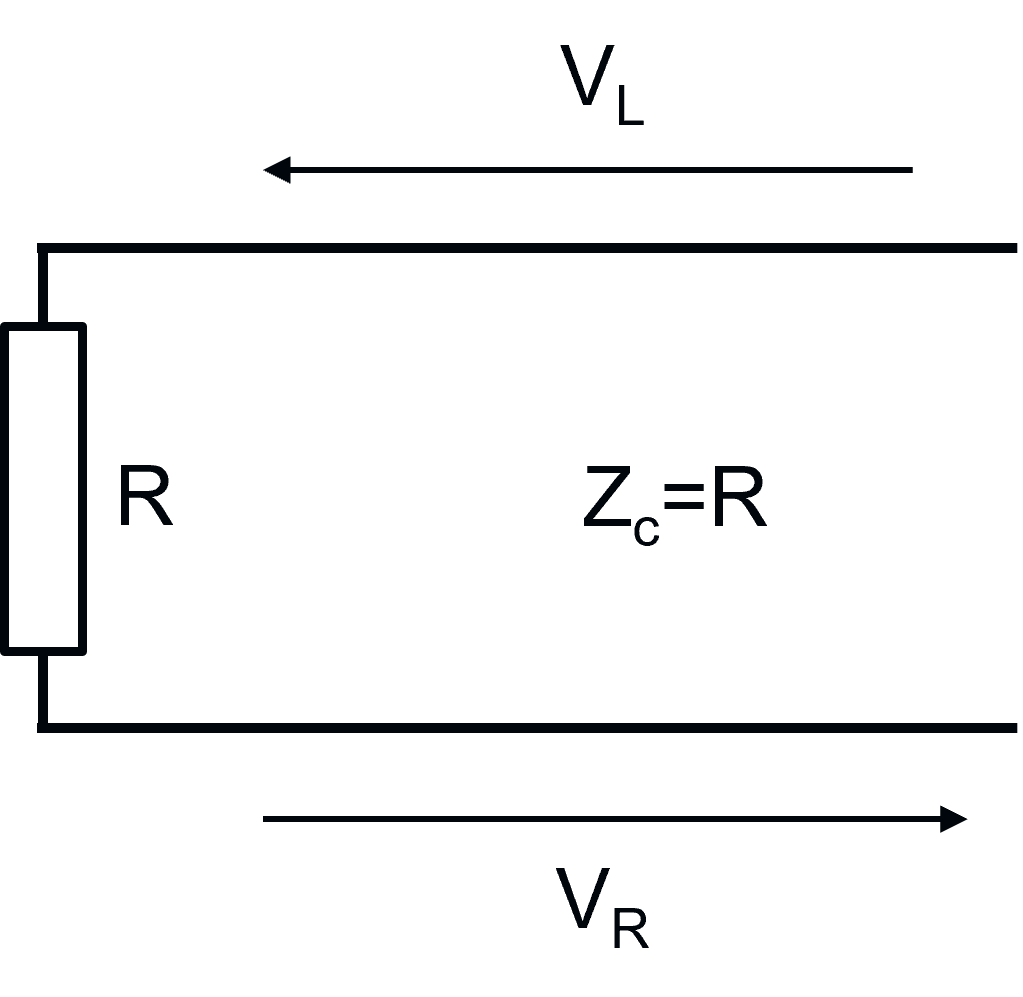}\\ 
\includegraphics[scale=0.5]{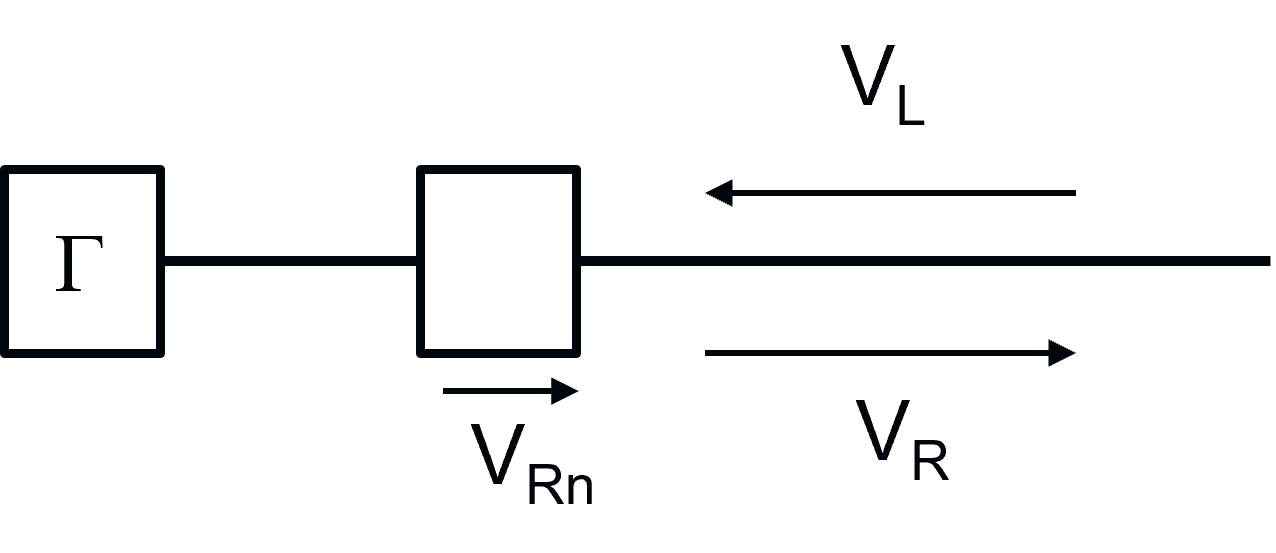}
\caption{{\bf 1-Port Thermal Noise.}
Panel (a) illustrates the classic Johnson-Nyquist
noise argument where the left-moving waves in a thermal state 
of the transmission line are
completely absorbed by the resistor, which in turn through its 
internal thermal fluctuations generates a right-moving wave 
such that the left-moving and right-moving power fluxes cancel.
In panel (b) this argument is generalized to an arbitrary 
1-port with a not necessarily vanishing reflection coefficient 
$\Gamma $ and an outgoing travelling wave noise source.
}
\label{fig_1port_thermal_noise}
\end{figure}

We first review and generalize the thermodynamic argument.
In the classic derivation of Johnson-Nyquist 
noise, a resistor of resistance $R,$
which may be understood as a 1-port, is connected to a 
transmission line with $Z_c=R.$ The TEM modes of the 
transmission line, which may be decomposed
into left-moving and right-moving modes, are assumed to be in a 
thermal state at a temperature $T$ equal to the temperature 
of the resistor. As indicated in
Fig.~\ref{fig_1port_thermal_noise}, 
the left-moving waves in this setup are completely absorbed by 
the resistor, which by design has been perfectly 
matched. In order to obtain zero net power transfer, 
the right-moving noise source must on the average
replenish the left-moving power flux at each frequency; 
consequently
$\left\langle \vert v_L\vert ^2 \right\rangle =R~k_BT(\Delta B).$
This argument generalizes to the case where $0<\vert \Gamma \vert \le 1,$
in which case we obtain
\begin{equation}
\left\langle \vert v_L\vert ^2 \right\rangle =
\left( 1-\vert \Gamma \vert ^2\right)
R~k_BT(\Delta B).
\end{equation}

\begin{figure}[H]%[!h]
\centering
\includegraphics[scale=0.4]{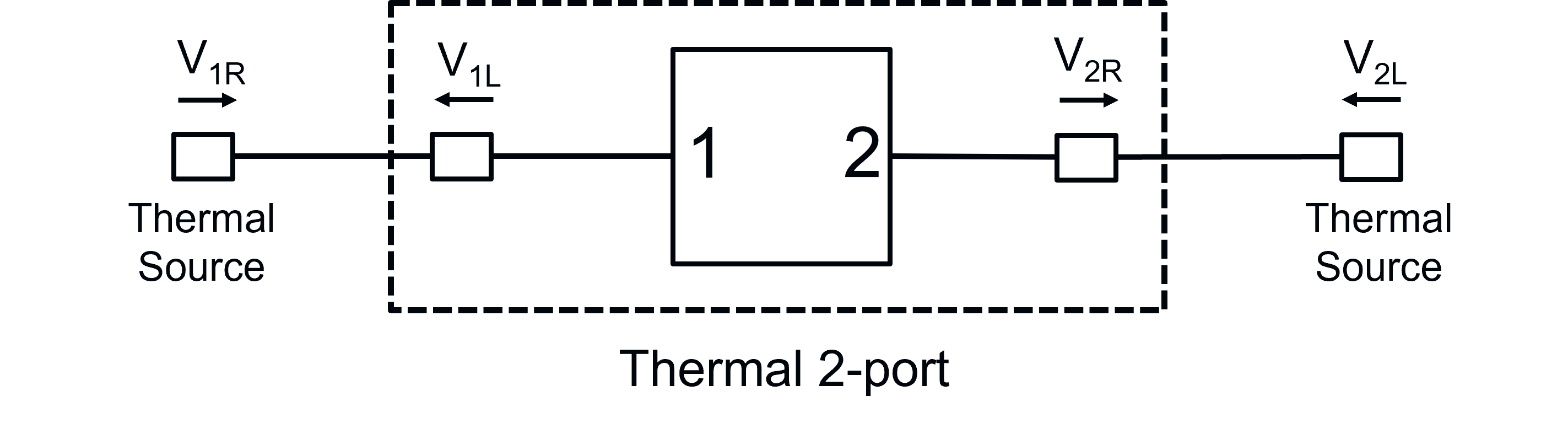}
\caption{{\bf 2-Port Thermal Noise.}
The outgoing travelling wave sources in the 2-port generate
the missing power absorbed within the thermal 2-port.
}
\label{fig_2port_thermal_noise}
\end{figure}

We now generalize to a thermal 2-port, as illustrated in Fig.~\ref{fig_2port_thermal_noise}.  Let us
for the moment turn off the outgoing travelling wave noise sources of the
two-port model $v_{1L}$ and $v_{2R}$ and calculate the power deficit of the
return power flux
assuming that the external noise sources $v_{1R}$ and $v_{2L}$ are
thermal sources. We obtain the following fractional power deficits
\begin{equation}
\begin{aligned}
\frac{
P_{1,return}
}{
P_{1,incident}
}
&= 1-\vert S_{11}\vert ^2- \vert S_{21}\vert ^2,\\
\frac{
P_{2,return}
}{
P_{2,incident}
}
&= 1-\vert S_{22}\vert ^2- \vert S_{12}\vert ^2.
\end{aligned}
\end{equation}
Consequently,
\begin{equation}
\begin{aligned}
\left\langle \vert v_{1L}\vert ^2 \right\rangle
&= \left( 1-\vert S_{11}\vert ^2- \vert S_{21}\vert ^2\right)
R~k_BT(\Delta B),\\
\left\langle \vert v_{2R}\vert ^2 \right\rangle
&= \left( 1-\vert S_{22}\vert ^2- \vert S_{12}\vert ^2\right)
R~k_BT(\Delta B).
\end{aligned}
\end{equation}
The same argument generalizes to describe the noise of a thermal $N$-port.

It may be verified that the expressions 
in eqns.~(\ref{eqn:CorrUnifTemp}) and (\ref{eqn:CorrUnifTempTW})
satisfy the thermodynamic
requirements derived in this section when all components are at the same temperature.

%\begin{figure}[H]%[!h]
%\centering
%\includegraphics[scale=0.35]{fig3_b.png}
%\caption{{\bf 1-Port Thermal Noise panel b).}
%In panel (b) this argument is generalized to an arbitrary 
%1-port with a not necessarily vanishing reflection coefficient 
%$\Gamma $ and an outgoing travelling wave noise source.
%}
%\label{fig_3_b}
%\end{figure}

\begin{figure}[H]%[!h]
\centering
\includegraphics[width=2.5in]{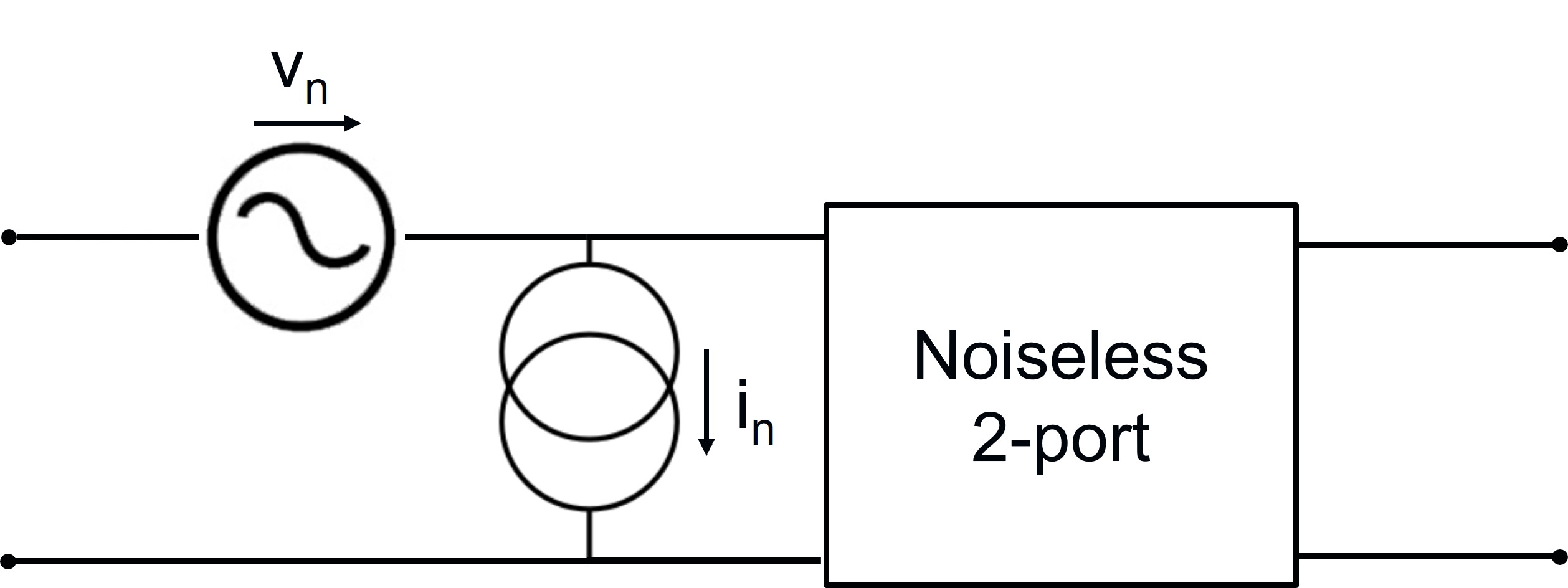}\\
(a)\\
\includegraphics[width=2.5in]{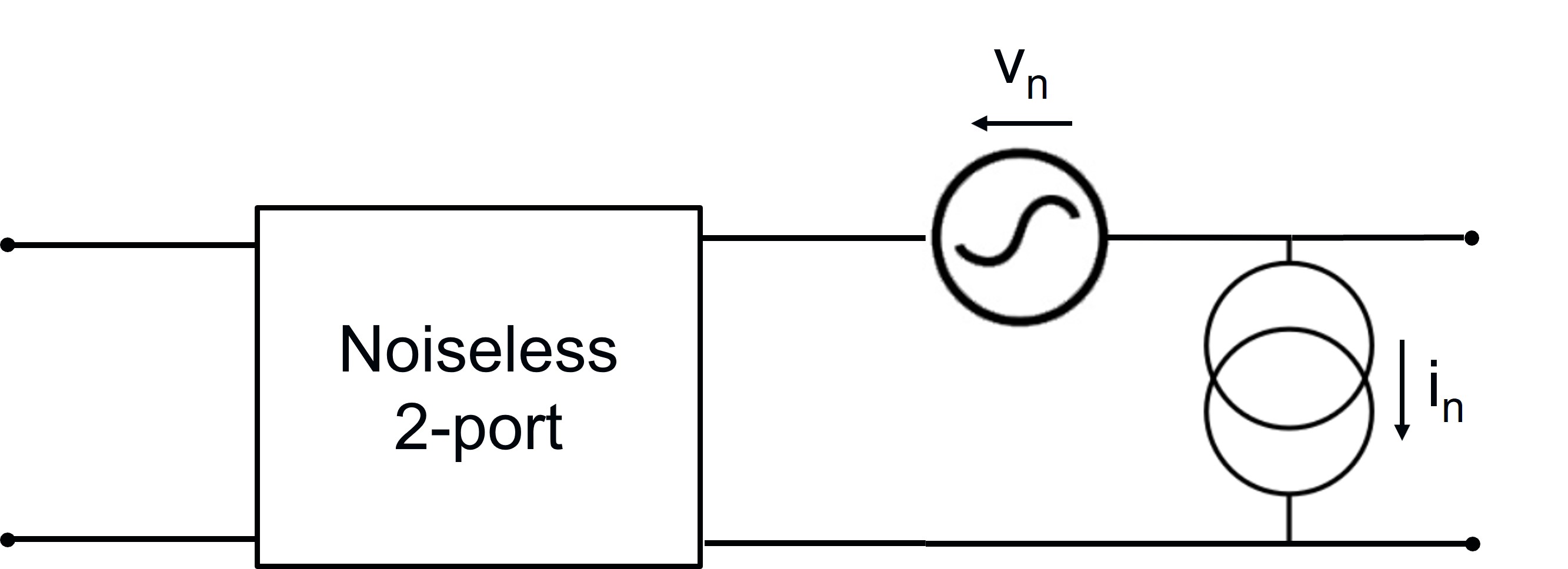}\\
(b)
\caption{Even though noise is generated everywhere along the cable, 
one can represent the noise generated within
the cable by an equivalent circuit consisting of a noiseless 2-port with current and 
voltage sources localized at either end of the cable, as illustrated in (a) and (b).
A noiseless cable is represented by a 2-port here.}
\label{fig_two}
\end{figure}
\section{Discussion}

The matrix 
\begin{equation}
\begin{pmatrix}
1 & \delta \mathcal{Z}\\
\delta \mathcal{G} & 1
\end{pmatrix}
\end{equation}
expresses the most general possible infinitesimal transfer matrix consistent with symmetry when
the ends are exchanged $(1\leftrightarrow 2)$ and reciprocity, which is equivalent
to requiring that the matrix have unit determinant. A linear perturbation to the diagonal
would violate reciprocity or reflection symmetry. We 
see that a four-parameter model follows from these very 
general considerations. This modelling is without approximation. 

In principle measurement of the reflection coefficient for two
different finite lengths of cable
with the same boundary condition, for example 
the pair
$\Gamma (L_1)^{(open)}$
and 
$\Gamma (L_2)^{(open)},$
or the pair
$\Gamma (L_1)^{(shorted)}$
and 
$\Gamma (L_2)^{(shorted)},$ suffices to determine 
$\mathcal{Z}$
and 
$\mathcal{Y}$
of the cable, as there are four real parameters to be
determined. Alternatively, one could measure the 2-port S 
parameters of a single length of cable. The S-parameters can be 
converted to T or Z matrices by applying the standard conversion 
formulas. These four real parameters completely determine
the noise properties assuming that the temperature profile along the 
cable is known. 

To the extent that the loss tangent depends weakly 
on frequency, the dielectric loss (in dB per unit
length) increases linearly with frequency (as opposed
to linearly with the square root of the frequency,
as is the case for the skin depth). 
Thus at higher frequencies dielectric loss dominates.
Teflon (R) [or 
PTFE (polytetrafluoroethylene)] has a low loss tangent
compared to other less expensive dielectrics suitable
at lower frequencies when dielectric loss is 
subdominant. 

%\begin{figure}[H]%[!h]
%\centering
%\includegraphics[width=2.5in]{dm_fig_one.pdf}
%\caption{Figure 1.}
%\label{fig_one}
%\end{figure}

%\begin{figure}[H]%[!h]
%\centering
%\includegraphics[width=2.5in]{dm_fig_three.pdf}
%\caption{Figure 3.}
%\label{fig_three}
%\end{figure}

%\begin{figure}[H]%[!h]
%\centering
%\includegraphics[width=2.5in]{dm_fig_four.pdf}
%\caption{Figure 4.}
%\label{fig_one}
%\end{figure}

%\section{Conclusion}

We have presented a general model of a lossy cable and the thermal noise generated within
in which the cable is represented by a lossless 2-port with a voltage and current noise 
sources on one end. We showed that in the case where the cable is at a uniform 
temperature the resulting noise properties are consistent with thermodynamic
expectations. However, the formula derived also extends to general case where 
the temperature profile along the cable is not uniform.

Whereas many models of lossy cables are parameterized by three real parameters, which
at a given fixed frequency could for example be taken to be a phase velocity, real 
characteristic impedance, and attenuation coefficient, our model includes four real 
parameters because the characteristic impedance in the lossy case is in general 
complex (with a non-vanishing imaginary part). These four parameters can be readily 
be measured, for example using a VNA, and these parameters together with the temperature 
profile completely determine the noise properties of a section of cable provided that
linearity holds and there are no DC bias currents.

Supplementary material with some examples implementing 
eqn.~(\ref{eqn:noiseCorr}) as well as a code may be found at:

\noindent
https://github.com/DannyMolnar/NoisyCable\_paper

\section*{Acknowledgments}

\noindent
MB acknowledges a SKA-LOFAR travel grant from the Observatoire de Paris
for a trip to Cambridge where part of this work was done. 
MB and DM thank Eloy de Lera Acedo, Dirk de Villiers, and especially 
Paul Scott for useful discussions and comments. 
MB and DM would like to thank the REACH collaboration for support 
and for inspiring this investigation.
DM also thanks Murray Edwards College at the University of Cambridge.


%\newpage

\begin{IEEEbiography}%
[{\includegraphics[width=1in,height=1.25in,clip,keepaspectratio]%
%{martinGibbon.jpg}}%
{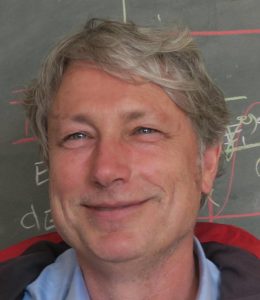}}]
{Martin Bucher}
is with the CNRS based at the
Laboratoire APC (Astroparticules et Cosmologie) at
Universit\'e Paris Cit\'e in Paris, France. Bucher
received his PhD in Physics from Caltech in 1990
and has held positions at the Institute of Advanced Study,
Princeton University, Stony Brook University, and DAMTP
at the University of Cambridge. Bucher held the SW Hawking
Fellowship of Mathematical Sciences at Trinity Hall, University
of Cambridge from 2000-2004 before coming to France where he
worked on the ESA Planck Mission, which mapped the microwave sky
in nine frequencies, thus constraining models of the primordial
universe. Bucher was recipient of the 2018 Gruber Prize 
in Cosmology
as part of the Planck team and is member of the Academy of 
Science of South Africa. Bucher is a part of the REACH 
collaboration. 
%When Bucher is not doing science, he can annoy 
%his colleagues by pulling their tails. 
\end{IEEEbiography}

\begin{IEEEbiography}%
[{\includegraphics[width=1in,height=1.25in,keepaspectratio]
{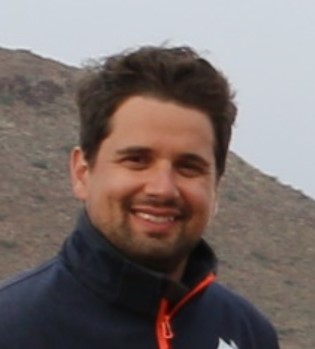}}]{Daniel Molnar}
%[{\includegraphics[width=1in,height=1.25in,keepaspectratio]
%{molnarPic.png}}]{Daniel Molnar}
is with the University of Cambridge. 
Daniel Molnar received an MPhil degree and
Ph.D.~in Physics from the University of
Cambridge, Cambridge, UK, in 2013 and 2016.
He has worked at the Department of Engineering
at the University of Cambridge, COMSOL Inc., and
the Theory of Condensed Matter group at the
Cavendish Laboratory, University of 
Cambridge. He later worked at CERN, the European
Particle Accelerator Laboratory, in Geneva, 
Switzerland as a Senior Fellow.
Dr. Molnar is a Bye-Fellow in Physics at Murray
Edwards College at the University of Cambridge.
Molnar is a member of the REACH Collaboration.
%When Daniel is not doing physics, he likes to hunt vegetarians.  
\end{IEEEbiography}

%MB add
%\begin{multicols}{2}
%\end{multicols}{2}
\end{multicols}
%\end{multicols}
\vfill

\end{document}